\documentclass{eas}
\usepackage{graphicx}
\begin{document}
\runningtitle{Hunt \& Kawata: M2M dynamical Milky Way modelling with Gaia}
\title{A PRIMAL view of the Milky Way, made possible by Gaia and M2M modelling} 
\author{Jason A. S. Hunt}\address{Mullard Space Science Laboratory, Holmbury St. Mary, Dorking, Surrey, RH5 6NT, UK}
\author{Daisuke Kawata}\sameaddress{1}
\begin{abstract}
We have developed our original made-to-measure (M2M) algorithm, \sc{primal}\rm, with the aim of modelling the Galactic disc from upcoming $Gaia$ data. From a Milky Way like $N$-body disc galaxy simulation, we have created mock $Gaia$ data using M0III stars as tracers, taking into account extinction and the expected Gaia errors. In \sc{primal}\rm, observables calculated from the $N$-body model are compared with the target stars, at the position of the target stars. Using \sc{primal}\rm, the masses of the $N$-body model particles are changed to reproduce the target mock data, and the gravitational potential is automatically adjusted by the changing mass of the model particles. We have also adopted a new resampling scheme for the model particles to keep the mass resolution of the $N$-body model relatively constant. We have applied \sc{primal }\rm to this mock Gaia data and we show that \sc{primal }\rm can recover the structure and kinematics of a Milky Way like barred spiral disc, along with the apparent bar structure and pattern speed of the bar despite the galactic extinction and the observational errors.
\end{abstract}
\maketitle
\section{Introduction}
\label{intro}

Understanding the structure and dynamics of the Milky Way is one of the oldest questions in astronomy. Despite centuries of effort, it remains hard to get a clear picture of the Galaxy we live in, owing to our position within the Galaxy, where much of the information is obscured by dust extinction, and complicated by observational errors. We need accurate survey data, and improved modelling techniques to understand and compensate for observational bias.

The European Space Agency's $Gaia$ mission, which was launched on $19^{th}$ December 2013, will provide positions and kinematics for around 1 billion stars, and is vastly superior to its predecessor $Hipparcos$ in terms of both quantity and quality of data collected. This new wealth of information about our Galaxy is invaluable to enhance our understanding of the Milky Way. However, only $\sim1\%$ of the stars in our Galaxy are bright enough to be observed by $Gaia$, and thus, to obtain a complete dynamical model of the Milky Way, we have to infer the structure of the Galaxy where $Gaia$ cannot see.

There are many different methods to construct dynamical galaxy models. However, most of them are restricted to use with axisymmetric systems, either by nature or by complexity. $N$-body methods are not restricted by symmetry, although creating a model that resembles target data is difficult due to the chaotic nature of $N$-body dynamics. The Made-to-Measure (M2M) method, pioneered by Syer \& Tremaine (\cite{ST96}) provides a way to tailor $N$-body models to match observations. It has been used to model the Milky Way (e.g. Bissantz $et$ $al.$ \cite{BDG04}) and external galaxies (e.g. de Lorenzi $et$ $al.$ \cite{deL08}) and construct initial conditions for $N$-body simulations (e.g. Dehnen \cite{Deh09}). The M2M method works by calculating observable properties from the model to be compared with the target data, and then adapting particle weights such that the model reproduces the target data. 

\section{\sc{primal}\rm: A self-consistent M2M algorithm for the $Gaia$ era}
We have presented a full description of our particle-by-particle M2M, \sc{primal }\rm, in Hunt \& Kawata (\cite{HK13}, \cite{HK14}) and Hunt $et$ $al.$ (\cite{HKM13}). In this section we describe briefly the basis of \sc{primal}\rm.

Traditional M2M algorithms work with test particles in a fixed or adaptive potential, and alter particle weights during the simulation. \sc{primal }\rm applies the M2M method to a live $N$-body model, altering the particle masses, which in turn alters the potential which is calculated self-consistently using the mass of the $N$-body particles. This naturally leads to structure formation, and allows us to more easily reproduce the non-axisymmetric structure, such as the bar and spiral arms, in the observed galaxy. As mentioned in Section \ref{intro}, our ultimate target is the Milky Way, where the observables are not binned data, but the position and velocity of the individual stars which are distributed rather randomly. Thus we compare the model and target observables at the position of each target star. 

With \sc{primal}\rm, we put constraints on the model based on the mass density and velocity of the observables in the target and the model. The observables are calculated using a spherically symmetric spline kernel often used in Smoothed Particle Hydrodynamics (SPH, Monaghan \& Lattanzio \cite{ML85}). For the density constraint, we use the kernel to calculate the density around each target star in both the target data, and the equivalent position in the model. For the velocity constraints, we maximise the likelihood of the three equatorial velocity components $(\mu_{\alpha},\mu_{\delta},v_r)$ of the target stars to be reproduced by the local velocity field of the model as shown in Hunt \& Kawata (\cite{HK14}). The likelihood function allows us to weight the contribution of the target stars by the error in each component of the velocity of each star.

In a new addition from Hunt \& Kawata (\cite{HK14}) we introduce the resampling of $N$-body model particles whose masses drift too far from the mean particle mass. We set a limit on how large or small these particle masses can become. We then delete any particle whose mass crosses the lower limit, $m_{\rm{min}}$ to save on computational time, and split any particle whose mass crosses the upper limit, $m_{\rm{max}}$ to prevent a single massive particle dominating local dynamics. The particles with $m_i>m_{\rm{max}}$ are split into the appropriate number of particles to keep their mass close to the mean particle mass, $\bar{m}$. The parent particle is retained with decreased mass, and generated particles are spaced randomly within the smoothing length of the kernel. All generated particles share all other properties with the parent particle including its velocity. 

For the resampling criteria in this example, we set $m_{\rm{max}}=3\bar{m}$ and $m_{\rm{min}}=\frac{1}{3}\bar{m}$, where $\bar{m}$ is the mean particle mass for particles which are closer than 10 kpc from the observer. This maintains a more even mass distribution around $\bar{m}$. We may replace the prior, $\hat{m}$, in the weight entropy regularization term in the M2M method with $\bar{m}$, resulting in regularization around a flexible prior, similar to the Moving Prior Regularization described in Morganti \& Gerhard (\cite{MG12}). Interestingly, we find that we can remove the weight entropy regularisation term and regularize the model with the resampling alone. Therefore we present the results of \sc{primal }\rm without the weight entropy regularization term, but with the resampling scheme above. This is a first attempt at running \sc{primal }\rm without the entropy term, and we have yet to explore which parameters and method of regularization works best with the resampling scheme.

\begin{figure}
\centering
\includegraphics[width=\hsize]{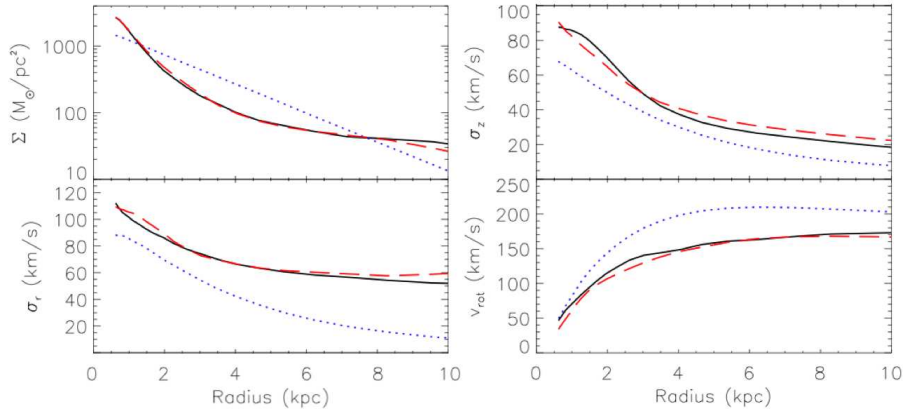}
\caption{Surface density profile (upper left), radial velocity dispersion (lower left), vertical velocity dispersion (upper right) and mean rotation velocity (lower right) for the initial model (blue dot), target galaxy (black solid) and the model by \sc{primal }\rm (red dash).}
\label{fig}
\end{figure}

\section{Results}
\label{R}
To test \sc{primal }\rm we created a target disc galaxy in a fixed dark matter potential using an $N$-body simulation. Our target data are M0III tracers from this $N$-body galaxy, assuming that each particle corresponds to one tracer. We use the 3D extinction map from Galaxia (Sharma $et$ $al.$ \cite{SBJB11}) to add extinction and calculate a $Gaia$-like error with a code provided by Merce Romero-G\'{o}mez (e.g. Romero-G\'{o}mez $et$ $al.$ \cite{MRG14}). To maintain a balance between quantity and quality of data our target stars are selected with $V\leq16.5$ mag, and the observed distance $d\leq10$ kpc. Fig. \ref{fig} shows the radial profiles of the surface density (upper left), radial velocity dispersion (lower left), vertical velocity dispersion (upper right) and mean rotation velocity (lower right) for the initial model (blue dot), target galaxy (black solid) and the model recovered by \sc{primal }\rm (red dash). The initial model was a smooth disc, and deliberately adopted a lower scale length than the target $N$-body galaxy. Fig. \ref{fig} shows that in each profile, \sc{primal }\rm has created a good reproduction of the target $N$-body galaxy. The surface density profile shows a particularly good recreation of the target galaxy, and the velocity profiles show a good recovery considering the error and extinction present in the target data. The target $N$-body galaxy had a bar pattern speed of $\Omega_{p,t} = 28.9$ km s$^{-1}$ kpc$^{-1}$ which \sc{primal }\rm recovers very well, resulting in $\Omega_p = 29.2$ km s$^{-1}$ kpc$^{-1}$.

\section{Summary \& further work}
We have shown that \sc{primal }\rm can make a dynamical galaxy model from mock $Gaia$ observations when using M0III tracers of the target $N$-body galaxy. When applying \sc{primal }\rm to the real $Gaia$ catalogue we will have to select tracer populations from the catalogue using astrophysical parameters, which will generate additional errors. Therefore to test and calibrate \sc{primal }\rm we have developed a population synthesis tool called \sc{snapdragons }\rm (Hunt $et$ $al.$ \cite{H15}) which can construct $Gaia$-like mock data from $N$-body models. These mock $Gaia$ observations will allow us to test our selection functions and determine which stellar types should be used as constraints for \sc{primal}\rm.


\end{document}